\documentclass[twocolumn,showpacs,%
  nofootinbib,aps,superscriptaddress,%
  eqsecnum,prd,notitlepage,showkeys,10pt]{revtex4-1}

\usepackage{amssymb}
\usepackage{amsmath}
\usepackage{graphicx}
\usepackage{dcolumn}
\usepackage{hyperref}
\usepackage{fancyhdr}
\usepackage{amsmath,color}
\usepackage{ragged2e}

\begin{document}

\title{Null Hypersurfaces in Kerr-Newman-AdS Black Hole and Super-Entropic Black Hole Spacetimes}
\author{Michael T.N. Imseis}
\email{mtimseis@uwaterloo.ca}
\author{Abdulrahim Al Balushi}
\email{abdulrahim.albalushi@uwaterloo.ca}
\author{Robert B. Mann}
\email{rbmann@uwaterloo.ca}
\affiliation{Department of Physics and Astronomy, University of Waterloo, Waterloo, Ontario, N2L 3G1, Canada}

\begin{abstract}
     A three-dimensional light-like foliation of a spacetime geometry is one particular way of studying its light cone structure and has important applications in numerical relativity. In this paper, we execute such a foliation for the Kerr-Newman-AdS black hole geometry  and compare it with the lightlike foliations of the Kerr-AdS and Kerr-Newman black holes. We derive the equations that govern this slicing and study their properties. In particular, we find that these null hypersurfaces develop caustics inside the inner horizon of the Kerr-Newman-AdS black hole, in strong contrast to the Kerr-AdS case. We then take the ultra-spinning limit of the Kerr-Newman-AdS spacetime, leading to what is known as a Super-Entropic black hole, and show that the null hypersurfaces develop caustics at a finite distance outside the event horizon of this black hole. As an application, we construct Kruskal coordinates for both the Kerr-Newman-AdS black hole and its ultra-spinning counterpart.
\end{abstract}

\maketitle

\section{Introduction}

One of the most important aspects of any spacetime geometry is its causal structure. The causal structure of a spacetime allows us to understand what cause and effect means within that geometry. Studying the causal structure of a spacetime is equivalent to studying how light cones behave within it. For example, light cones in Minkowski spacetime are spherical and propagate outward with an areal radius that is affinely parameterized. Thus, Minkowski spacetime has a well defined notion of causality. Generalizations of this to spherically symmetric black holes can be straightforwardly obtained.\par

Studying the light cone structure of more complicated geometries presents more of a challenge. 
Three-dimensional null foliations have already been studied for more complicated geometries such as the Kerr black hole \cite{pretorius1998quasi} and, only recently \cite{al2019null}, the Kerr-AdS black hole. In particular, it was shown that null hypersurfaces in Kerr geometries were free of caustics for all $r>0$ and asymptotically approached Minkowski space light cones as $r\rightarrow \infty$ \cite{pretorius1998quasi}. The light cone structure of rotating black holes embedded in AdS was then studied, and it was shown that once again the light cones possessed no caustics for $r>0$ and asymptotically approached pure AdS light cones as $r\rightarrow \infty$ \cite{al2019null}. 

Here we investigate   what happens to these light cones when the Kerr-AdS black hole is electrically charged, namely for  the Kerr-Newman-AdS metric \cite{Carter:1968ks}.
We study the structure of null hypersurfaces in Kerr-Newman-AdS spacetimes (KNAdS) by introducing a three-dimensional lightlike slicing of this geometry, and develop the equations governing such a slicing. We then take the ultra-spinning limit of the Kerr-Newman-AdS metric and study how this effects the foliation.

Our key result of this paper is the fact that caustics form inside the inner horizon of the KNAdS black hole, in stark contrast to the uncharged Kerr-AdS case \cite{al2019null}. Caustics also form at a finite distance \textit{outside} the black hole when the ultra-spinning limit is taken. This can be attributed to the ZAMO angular velocity of the ultra-spinning geometry, as will be discussed in later sections.

Our paper is organized as follows. We begin in section II by writing down the equations governing the null foliation and solve them in terms of elliptic integrals. Section III will focus on studying the properties of the null hypersurfaces in a KNAdS geometry. In section IV, we provide a proof that caustics form inside the inner horizon of the KNAdS black hole. Section V will be focused on taking the ultra-spinning limit of the KNAdS geometry and proving that caustics form at a finite distance outside the black hole, when this limit is taken. Finally, as an application, the construction of Kruskal coordinates for the KNAdS black hole and its ultra-spinning counterpart will be outlined in section VI.

%\\\\\\\ SECTION 2 \\\\\\\

\section{Equations of Null Hypersurfaces}
To begin, the 4-dimensional Kerr-Newman-AdS metric \cite{carter1968hamilton,plebanski1976rotating}, written in standard Boyer-Lindquist form \cite{hawking1999rotation}, which describes an asymptotically AdS, rotating, electrically charged black hole is
\begin{widetext}
\begin{equation}\label{KNAdS}
    ds^2 = -\frac{\Delta_{r}}{\Sigma^2}\left(dt-\frac{a}{\Xi}\sin^2\theta\: d\phi \right)^2\:+\frac{\Sigma^2}{\Delta_{r}}dr^2\:+\frac{\Sigma^2}{\Delta_{\theta}}d\theta^2+\frac{\Delta_{\theta}}{\Sigma^2}\sin^2\theta\:\left(a\:dt\:-\frac{r^2+a^2}{\Xi}d\phi\right)^2
\end{equation}
where, 
\begin{equation}\label{KNAdS2}
    \Delta_{r}\:=(r^2+a^2)\left(1+\frac{r^2}{l^2}\right)-2mr+q^{2}, \hspace{0.5cm} \Xi\:=1-\frac{a^2}{l^2}, \hspace{0.5cm}
    \Delta_{\theta}\:=1-\frac{a^2\cos^2\theta}{l^2},\hspace{0.5cm} \Sigma^2\:=r^2\:+a^2\cos^2\theta
\end{equation}    
\end{widetext}

In this metric, $a$ is the rotation parameter of the black hole, $l$ is the AdS length and $q$ is the electric charge of the black hole. Horizons of this black hole are defined through roots of the metric function $\Delta_{r}$, namely $\Delta_{r}(r_{\pm})=0$ for the inner and outer horizons, $r_{\pm}$, respectively. Here, we assume that $a<l$ so that the outer horizon $r_{+}$ is given by the largest root of $\Delta_{r}$. The metric \eqref{KNAdS} is written in terms of a coordinate system that is rotating at infinity with an angular velocity of $\Omega_{\infty} = -a/l^{2}$. Also, the azimuthal coordinate $\phi$ is a compact coordinate with range $[0,2\pi]$.

The relevant thermodynamic parameters for the KNAdS black hole are given by \cite{caldarelli2000thermodynamics,Anabalon:2018qfv}
\begin{gather}
    M=\frac{m}{\Xi^{2}}, \hspace{0.3cm} J = \frac{am}{\Xi^{2}}, \hspace{0.3cm} \Omega = \frac{a(1+r_{+}^{2}/l^{2})}{r_{+}^{2}+a^{2}}, \hspace{0.3cm} \notag \\ 
    A = \frac{4\pi\left(r_{+}^{2}+a^{2}\right)}{\Xi}, \hspace{0.3cm} S = \frac{A}{4}, \hspace{0.3cm} Q_{e} = \frac{q}{\Xi},  \\
    \Phi_{e} = \frac{\pi Q_{e}}{2MS}\left(Q_{e}^{2} + \frac{S}{\pi} + \frac{S^{2}}{\pi^{2}l^{2}}\right), \notag \\
    T = \frac{1}{8\pi M}\left[1 - \frac{\pi^{2}}{S^{2}}(4J^{2}+Q_{e}^{4}) + \frac{2}{l^{2}}\left(Q_{e}^{2} + \frac{2S}{\pi}\right) + \frac{3S^{2}}{\pi^{2}l^{4}}\right] \notag \\
   \textsf{P} = \frac{3}{8\pi l^2} \qquad
    V= \frac{4\pi}{3}\left[r_+(r_{+}^{2}+a^{2})+\frac{m a^2}{\Xi} \right] \notag
\end{gather}
where $M$, $J$, $\Omega$, $A$, $S$, $Q_{e}$, $\Phi_{e}$, and $T$, $\textsf{P}$, $V$ are the mass, angular momentum, angular velocity, area, entropy, electric charge, electric potential,  temperature,
pressure, and thermodynamic volume respectively. Here, $M$ and $J$ have been calculated by means of Komar integrals using the killing vectors $\partial_{t}/\Xi$ and $\partial_{\phi}$ respectively \cite{caldarelli2000thermodynamics}.  The thermodynamic parameters in (2.3) satisfy the first law of black hole thermodynamics
\begin{gather}
    dM = TdS + \textsf{P} dV + \Omega dJ + \Phi_{e} dQ_{e}
\end{gather}
the Smarr relation \cite{Anabalon:2018qfv}
\begin{equation}
    M = 2(T S + \Omega J - \textsf{P} V ) + \Phi_e Q_e 
\end{equation}
and the  Christodoulou-Ruﬃni mass relation
\begin{gather}
    M^{2} = \frac{A}{16\pi} + \frac{\pi}{A}\left(4J^{2}+Q_{e}^{4}\right)+\frac{Q_{e}^{2}}{2} + \frac{J^{2}}{l^{2}} \nonumber\\
     + \frac{A}{8\pi l^{2}}\left(Q_{e}^{2}+\frac{A}{4\pi}+\frac{A^{2}}{32\pi^{2}l^{2}}\right)
\end{gather}
sometimes referred to as the
 generalized Smarr formula  \cite{caldarelli2000thermodynamics}.
 
%\abd{although it may be annoying, it is better to not have Q for both charge and in \eqref{PQdef}. For example, change electric charge to $Q_e$ and electric potential to $\Phi_e$ for minimum pain.}

%\rbm{Have we checked that 1st and Smarr hold for these variables?}

Null hypersurfaces are defined by some function $\Phi(x) = const$. The goal is to construct equations of this form for the spacetime \eqref{KNAdS}. We start by defining the ingoing and outgoing Eddington-Finkelstein coordinates 
\begin{align}\label{uvdef}
v=t+r_{*} \hspace{0.5cm} u=t-r_{*}
\end{align}
where $r_{*}$ is the so called \textit{tortoise} coordinate. The null hypersurfaces will be described by $v=v_{0}$ and $u=u_{0}$, where $(v_{0}, u_{0})$ are both constants. These are respectively called the ingoing and outgoing null generators of the hypersurfaces. The condition that the hypersurfaces be null is expressed as 
\begin{align}\label{2.5}
    g^{\mu\nu}\partial_{\mu}v\:\partial_{\nu}v=g^{tt}+g^{rr}(\partial_{r}r_{*})^2+g^{\theta\theta}(\partial_{\theta}r_{*})^2=0
\end{align}
It should be noted that we would have obtained the same equation had we used $u$ rather than $v$. The idea is to figure out what $r_{*}(r,\theta)$ needs to be  in order to describe null hypersurfaces in the background geometry \eqref{KNAdS}. This amounts to solving a partial differential equation. Once $r_{*}(r,\theta)$ is found, insertion of the result into \eqref{uvdef} yields the equations for null hypersurfaces for the KNAdS metric \eqref{KNAdS}.

Noting that
\begin{equation} \label{2.8}
    g^{tt} = \frac{g_{\phi\phi}}{g_{tt}g_{\phi\phi}-g_{t\phi}^{2}} = \frac{\Xi^{2}\left[a^{2}\Delta_{r}\sin^{2}\theta - \Delta_{\theta}(r^{2}+a^{2})^{2}\right]}{\Delta_{r}\Delta_{\theta}\Sigma^{2}}
\end{equation}
the partial differential equation that must be solved is 
\begin{equation}\label{rstareq}
    \Delta_{r}\left(\partial_{r}r_{*}\right)^{2} + \Delta_{\theta}\left(\partial_{\theta}r_{*}\right)^{2} = \Xi^{2}\left[\frac{(r^{2}+a^{2})^{2}}{\Delta_{r}} - \frac{a^{2}\sin^{2}\theta}{\Delta_{\theta}}\right]
\end{equation}
which has a separable form. In order to solve this PDE for $r_{*}(r,\theta)$, we define two new functions
\begin{gather}
    Q^{2}(r) = \Xi^{2}\left[(r^{2}+a^{2})^{2} - a^{2}\lambda \Delta_{r}\right], \notag \\ 
    P^{2}(\theta) = \Xi^{2}a^{2}\left[\lambda\Delta_{\theta} - \sin^{2}\theta\right]
\label{PQdef}    
\end{gather}
where $\lambda < 1$ is the separation constant  \cite{pretorius1998quasi,al2019null}. 

Having defined these new functions \eqref{PQdef}, the ansatz 
\begin{align}\label{2.6}
    \partial_{r}r_{*} = \frac{Q}{\Delta_{r}}, \hspace{0.5cm} \partial_{\theta}r_{*} = \frac{P}{\Delta_{\theta}} 
\end{align}
 can be shown that these satisfy \eqref{rstareq}. A solution to \eqref{rstareq} is then obtained by integrating the exact differential
\begin{equation}\label{dreq}
    dr_{*} = \frac{Q}{\Delta_{r}}dr + \frac{P}{\Delta_{\theta}}d\theta
\end{equation}
There will be an integration constant once \eqref{dreq} is integrated which will be denoted $a^{2}g(\lambda)/2$, where $a$ is the rotation parameter of the KNAdS black hole. 

To find a more general solution to \eqref{rstareq}, we follow the procedure in \cite{pretorius1998quasi},   promoting $\lambda$ to be a variable, yielding $r_{*} = \rho(r,\theta,\lambda)$. We then obtain
\begin{equation}\label{drhoeq}
    d\rho = \frac{Q}{\Delta_{r}}dr + \frac{P}{\Delta_{\theta}}d\theta + \frac{a^{2}}{2}Fd\lambda
\end{equation}
where $\partial_{\lambda}\rho = \frac{a^{2}}{2}F$ and 
\begin{equation}\label{2.12}
    F(r,\theta,\lambda,m) = \int_{r}^{\infty}\frac{dr^{\prime}}{Q(r^{\prime},\lambda)} + \int_{0}^{\theta}\frac{d\theta^{\prime}}{P(\theta^{\prime},\lambda)} +g^{\prime}(\lambda)
\end{equation}
For \eqref{dreq} and \eqref{drhoeq} to be functionally equivalent, the condition
\begin{equation}\label{Feq0}
    F=0
\end{equation}
needs to hold. This condition is going to fix the dependence of $\lambda$ strictly in terms of $r$ and $\theta$, for any given choice of the function $g(\lambda)$. Therefore, the most general solution to \eqref{rstareq} is 
%\rbm{Is $a$ the $a$ in the KAdS metric?}
\begin{equation}\label{2.14}
    r_{*} = \rho(r,\theta,\lambda(r,\theta)) = \int_{0}^{r} \frac{Q}{\Delta_{r}}dr + \int_{0}^{\theta} \frac{P}{\Delta_{\theta}}d\theta + \frac{a^{2}}{2}g(\lambda)
\end{equation}
Once a choice for $g(\lambda)$ is made, the integrals in \eqref{2.14} are done with $\lambda$ assumed to be constant. The constraint \eqref{Feq0} is then used to find the functional dependence of $\lambda$ on $r$ and $\theta$, which is then substituted back into the results from the integrals in \eqref{2.14}. Once this is complete, $r_{*} = \rho(r,\theta,\lambda)$ is obtained and the null generators in \eqref{uvdef} are found. It is possible to express each of the integrals \eqref{2.14} in terms of elliptic integrals.

%\\\\\\\ SECTION 3 \\\\\\\

\section{Properties of Null Hypersurfaces in $\text{KNAdS}$ Geometries}

The condition from \eqref{Feq0} implies that $dF=0$. This yields
\begin{equation}\label{mdl}
    \mu d\lambda = -\frac{dr}{Q} + \frac{d\theta}{P}
\end{equation}
\noindent where $\mu \equiv -\partial_{\lambda}F$. It then follows from \eqref{dreq} and \eqref{mdl} that ${\nabla r_*} \cdot \nabla \lambda = 0$ with respect to the intrinsic 2-metric 
\begin{equation}\label{sigmet}
     d\sigma^{2}=\frac{\Sigma^{2}}{\Delta_{r}}dr^{2} + \frac{\Sigma^{2}}{\Delta_{\theta}}d\theta^{2}
\end{equation}
 
\begin{figure}
    \centering
    \includegraphics[scale=0.30]{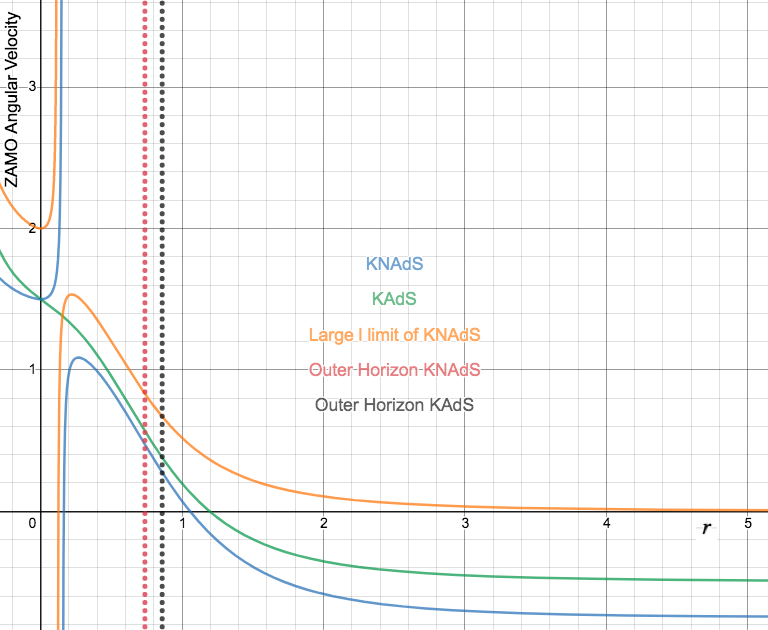} 
    \caption{The blue line shows the ZAMO angular velocity of inertial frame dragging in a Kerr-Newman-AdS geometry for $m=1$,\hspace{0.15cm} $a=0.5$,\hspace{0.15cm} $l=1$, \hspace{0.15cm} $q=0.5$ and $\theta=\pi/2$. The green line shows the ZAMO angular velocity in a Kerr-AdS geometry for the same parameters. As $q$ increases the plot shifts downward and as
    $l$ increases, the plot gets shifted upward.
    The yellow line is the $l\rightarrow\infty$ limit of $\omega$ for KNAdS, keeping the other parameters fixed.}
%\rbm{Do you agree with my statement about $q$?}    }
    \label{fig:zamo_AV}

\end{figure}

\noindent This proves that $r_{*}$ and $\lambda$ are orthogonal to one another on surfaces of constant $(t, \phi)$. It can also be shown that $\lambda$ is constant along the null generators defined in \eqref{uvdef}, through the equations
\begin{align}
    \ell^{\alpha}\partial_{\alpha}\lambda=0 \qquad
    n^{\alpha}\partial_{\alpha}\lambda=0
\label{ell-n}    
\end{align}
where $\ell_{\alpha}=-\partial_{\alpha}v$ and
$n_{\alpha}=-\partial_{\alpha}u$. This follows from the fact that $\lambda$ is independent of the coordinates $t$ and $\phi$. Making use of \eqref{dreq} and \eqref{mdl}, we can rewrite the intrinsic 2-metric \eqref{sigmet} as 
\begin{equation}\label{3.4}
    d\sigma^{2} = \frac{1}{\Xi^{4}R^{2}}\left[\Delta_{r}\Delta_{\theta} dr_{*}^{2} + \mu^{2}P^{2}Q^{2}d\lambda^{2}\right]
\end{equation}
\noindent where 
\begin{equation}\label{3.5}
    R^{2} \equiv \frac{g_{\phi\phi}}{\sin^{2}\theta} = \frac{\Delta_{\theta}(r^{2}+a^{2})^{2} - \Delta_{r}a^{2}\sin^{2}\theta}{\Xi^{2}\Sigma^{2}}
\end{equation}
\noindent We now rearrange \eqref{2.8} to give
\begin{equation}
    g_{tt} - \omega^{2}g_{\phi\phi} = \frac{1}{g^{tt}}
\end{equation}
yielding 
\begin{equation}
    \frac{\Delta_{r} - a^{2}\Delta_{\theta}\sin^{2}\theta}{\Sigma^{2}} + \omega^{2}R^{2}\sin^{2}\theta = \frac{\Delta_{r}\Delta_{\theta}}{\Xi^{4}R^{2}}
\end{equation}
where
\begin{equation}\label{3.8}
    \omega \equiv \frac{-g_{t\phi}}{g_{\phi\phi}} = \frac{a[\Delta_{\theta}(r^{2}+a^{2}) - \Delta_{r}]}{\Xi R^{2}\Sigma^{2}}
%\vspace{0.5cm}
\end{equation}
is the angular velocity of inertial frame dragging, due to the rotation of the black hole. The blue line in figure \ref{fig:zamo_AV} is a plot of this angular velocity as a function of distance from the KNAdS black hole. The most peculiar feature of this plot is that when $r$ is small, $\omega >0$. This means that when an inertial observer is very close to the black hole, spacetime is being dragged in the same direction as the black hole is rotating. However, as $r$ increases, $\omega$ becomes negative. This is signalling that as the observer gets farther away from the black hole, spacetime is being dragged in the {\it opposite} direction with respect to how the black hole is rotating. This is also true for the KAdS black hole, as shown by the green line in figure 1. In both cases, this peculiar feature of $\omega$ is due to the fact that we are working in a coordinate system that is rotating at infinity and is not an intrinsic feature of KNAdS or KAdS black holes, in general. 
%\rbm{We should check as to whether or not this is a generally well-understood feature of KAdS. How does the charge affect this property?}
In fact, one can also see from the plot that as $r$ increases, $\omega$ approaches a constant value in both cases. It can be shown that this constant value is indeed $\Omega_{\infty}=-a/l^{2}$ which is the angular velocity of the coordinate system at infinity. Here, the key difference between the KNAdS and KAdS black holes is that $\omega$ diverges inside the KNAdS black hole, while it remains smooth for $r>0$ in the KAdS case, as can be seen from the plot. This is due to the presence of $q\neq 0$ in \eqref{3.5} and \eqref{3.8}, for the KNAdS black hole. One interesting thing to note is that the divergence in $\omega$ arises due to the fact that $R$ from \eqref{3.5} has a root at $r>0$ inside the inner horizon of the black hole for any nonzero $q$. This is precisely the point at which closed timelike curves begin to form around the ring singularity of the KNAdS black hole. This is in contrast to the KAdS case where CTC's only form for $r<0$. Note also that the intrinsic 2-metric \eqref{3.4} becomes singular here.
\par
%\rbm{From what I can see in the plot, $\omega$ is continuous, but not differentiable at two points inside the outer horizon.  I have changed the text.  But why does this happen?}
\newpage
The yellow line in figure \ref{fig:zamo_AV} is a plot of $\omega$ for the KNAdS black hole, in the limit where $l\rightarrow \infty$. As the AdS length increases, the plot gets shifted upwards. The yellow line is in 
perfect agreement with the behaviour of $\omega$ for an asymptotically flat Kerr-Newman black hole, which makes sense because the $l\rightarrow \infty$ limit of an AdS spacetime is a flat spacetime. Note that in the $l\rightarrow\infty$ limit, $\omega > 0$ for all values of $r$, indicating that the inertial frames are always dragged in the same direction as the black hole, with respect to a reference frame at infinity. However, the divergence of $\omega$ inside the black hole remains present. 

Making use of the equations \eqref{3.4} - \eqref{3.8}, we can express the full KNAdS metric \eqref{KNAdS} in terms of the coordinates $(t, r_{*}, \lambda, \phi)$. Doing this yields
\begin{align}
    ds^{2} &= \frac{\Delta_{r}\Delta_{\theta}}{\Xi^{4}R^{2}}\left(dr_{*}^{2} - dt^{2}\right) + R^{2}\sin^{2}\theta\left(d\phi - \omega dt\right)^{2} \nonumber\\ 
    &\qquad + \frac{P^{2}Q^{2}\mu^{2}}{\Xi^{4}R^{2}}d\lambda^{2}
\label{3.9}    
\end{align}
for the KNAdS metric. This form of the metric has distinct advantages over
\eqref{KNAdS}: since the ingoing and outgoing null generators, defined in \eqref{uvdef}, are constant ($dv=du=0$), then $dr_{*}^{2} = dt^{2}$. Inserting  this result into \eqref{3.9}  gives
\begin{equation}\label{ind_met}
    dh^{2} = \frac{P^{2}Q^{2}\mu^{2}}{\Xi^{4}R^{2}}d\lambda^{2} + R^{2}\sin^{2}\theta\left(d\phi - \omega dt\right)^{2}
\end{equation}
for  the induced metric $dh^{2}$ on a light cone
in a KNAdS background. The form of this metric shows that the null generators are themselves rotating with the ZAMO angular velocity $\omega$, relative to observers at infinity. This can also be seen directly from the fact that $\ell_{\phi} = -\partial_{\phi}v=0$ and $n_{\phi} = -\partial_{\phi}u=0$. One should note that inversion of the differentials \eqref{dreq} and \eqref{mdl} are
\begin{align}
    dr = \frac{Q\Delta_{r}}{\Xi^{4}R^{2}\Sigma^{2}}\left[\Delta_{\theta}dr_{*} - P^{2}\mu d\lambda\right] \notag \\
    d\theta = \frac{P\Delta_{\theta}}{\Xi^{4}R^{2}\Sigma^{2}}\left[\Delta_{r}dr_{*} + Q^{2}\mu d\lambda \right]
\label{3.11}    
\end{align}
The equations of null hypersurfaces in KNAdS black hole geometries are respectively given by $v = t+r_{*}(r, \theta) = v_{0}$ and $u = t-r_{*}(r, \theta) = u_{0}$ where $(v_{0}, u_{0})$ are constants and $r_{*} = \rho(r, \theta, \lambda(r, \theta))$ where the function $\lambda(r, \theta)$ is determined by the constraint \eqref{Feq0}, analogous to \cite{pretorius1998quasi,al2019null}.

%\\\\\\\ SECTION 4 \\\\\\\
\section{Formation of Caustics for $0<r<r_{-}$}

From \eqref{ind_met}, the condition for caustic formation is 
\begin{equation}\label{4.1}
    PQ\mu\sin\theta \rightarrow 0
\end{equation}
where $\mu$ is defined in \eqref{mdl}. This is obtained by taking the determinant of \eqref{ind_met}. Since the volume element of a hypersurface is the square root of the determinant of the induced metric on that hypersurface, if the determinant goes to zero anywhere in the spacetime, the volume form will also go to zero, signalling the formation of a caustic. We proceed by analyzing each factor in \eqref{4.1} along an ingoing null generator with $\lambda = constant$ and $r$ decreasing. 

 We begin by noting that from \eqref{mdl}, $r$ is decreasing when $\theta$ decreases, for fixed $\lambda$. This means that $P>0$ \textit{increases} as $r$ decreases, along an ingoing generator when $\lambda$ is fixed. The next thing we show is that $\theta$ and $\mu$ both increase with increasing $m$. Note that from \eqref{3.11}, $\theta$ can be written as $\theta(r,\lambda,m)$.

\begin{figure}
   \centering
    \includegraphics[scale=0.29]{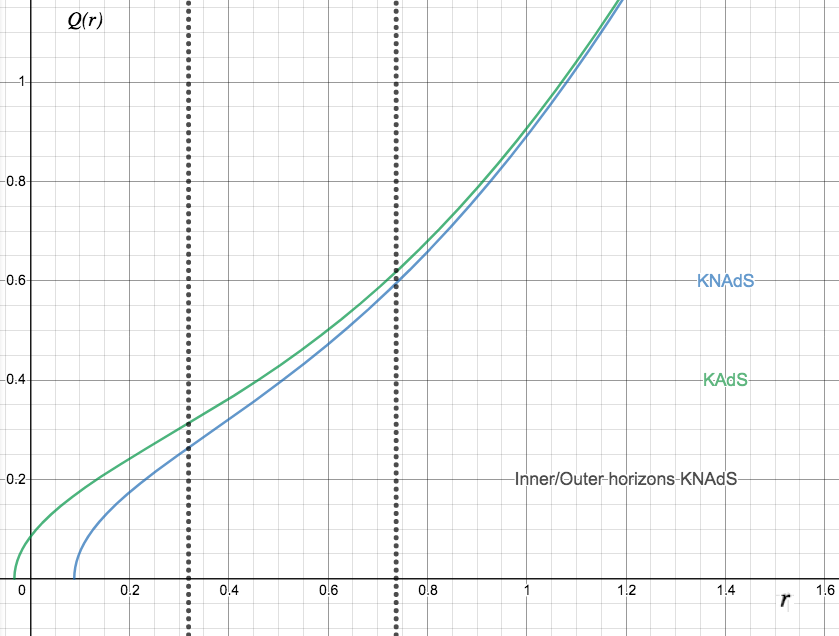}
    \caption{$Q(r)$ for the Kerr-AdS (green) and Kerr-Newman-AdS (blue) cases for $m=1$,\hspace{0.15cm} $a=0.5$,\hspace{0.15cm} $l=1$, \hspace{0.15cm} $q=0.5$, \hspace{0.15cm} $\lambda = 0.8$. The two dotted black lines are the locations of the inner and outer horizons, $(r_{-}, r_{+})$, of the KNAdS black hole respectively. For these particular parameters, $q=0.25$ is the threshold charge in order for caustics to develop, in the KNAdS case.}
    \label{fig:caustics_k(n)ads}
\end{figure}
\begin{figure}
    \centering
    \includegraphics[scale=0.29]{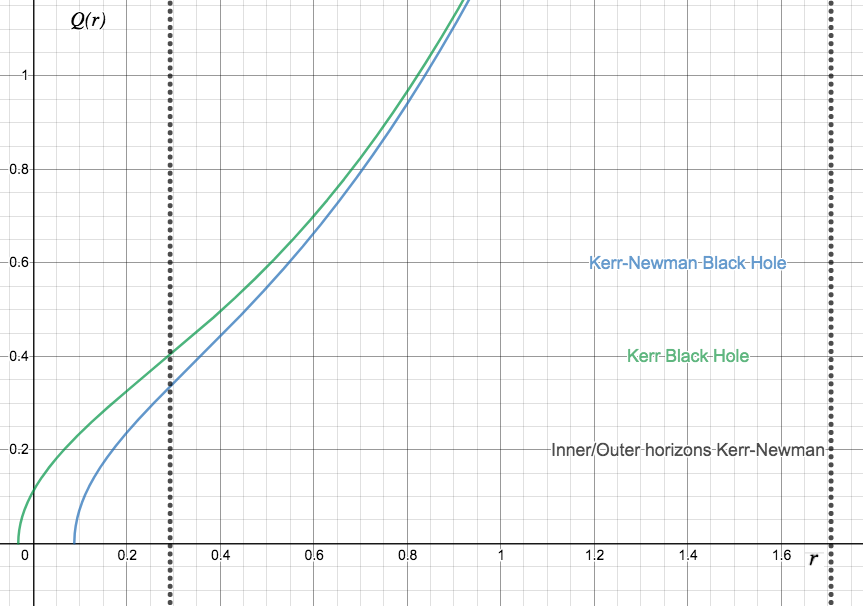}
    \caption{$Q(r)$ in the $l\rightarrow\infty$ limit of KAdS and KNAdS, otherwise known as Kerr and Kerr-Newman Black holes, for $m=1$,\hspace{0.15cm} $a=0.5$,\hspace{0.15cm} $q=0.5$,\hspace{0.15cm} $\lambda = 0.8$. The two dotted black lines are the locations of the inner and outer horizons, $(r_{-}, r_{+})$, of the KN black hole respectively. Caustics will appear for $q>0.25$.}
    \label{fig:caustics_l_infty}
 \end{figure}

Taking the differential of $F(r,\theta,\lambda,m)$ defined in \eqref{2.12} keeping $r$ and $\lambda$ fixed, setting $g(\lambda)=0$ without any loss of generality, and using condition \eqref{Feq0} yields
\begin{align}
    dF = 0 = \frac{\partial F}{\partial \theta} d\theta + \frac{\partial F}{\partial m} dm
\end{align}
From here, it is straightforward to show using \eqref{2.12}
\begin{align}
    \left(\frac{\partial\theta}{\partial m}\right)_{r,\lambda} = \Xi^{2}a^{2}\lambda P\int_{r}^{\infty}\frac{r^{\prime}dr^{\prime}}{Q^{3}}
\end{align}
which is manifestly positive for $r>0$. This shows that as the mass of the black hole increases, so does $\theta$. Now turning to $\mu$, we have
\begin{align}
    \left(\frac{\partial\mu}{\partial m}\right)_{r,\lambda} = \frac{\partial\mu}{\partial m} + \frac{\partial\mu}{\partial\theta}\left(\frac{\partial\theta}{\partial m}\right)_{r,\lambda}
\end{align}
\noindent Using the fact that $\mu \equiv -\partial_{\lambda} F$ and \eqref{2.12}, we obtain
\begin{align}
    \frac{\partial\mu}{\partial m} = -\frac{\partial}{\partial \lambda}\frac{\partial F}{\partial m} = \Xi^{2}a^{2}\frac{\partial}{\partial\lambda}\left(\lambda I\right)
\end{align}
where
\begin{align}\label{4.6}
    I \equiv \int_{r}^{\infty} \frac{r^{\prime}dr^{\prime}}{Q^{3}}
\end{align}
Next,
\begin{align}
    \frac{\partial\mu}{\partial\theta} = -\frac{\partial}{\partial\theta}\frac{\partial F}{\partial\lambda} = -\frac{\partial}{\partial\lambda}\frac{\partial F}{\partial\theta} = -\frac{\partial}{\partial\lambda}\left(\frac{1}{P}\right)
\end{align}
which gives 
\begin{align}
    \frac{\partial\mu}{\partial\theta} = \frac{\Xi^{2}a^{2}\Delta_{\theta}}{2P^{3}}
\end{align}

\noindent Note also that
\begin{align}\label{4.9}
    J \equiv \frac{\partial I}{\partial\lambda} = \int_{r}^{\infty}\frac{r^{\prime}\Delta_{r}}{Q^{5}}dr^{\prime}
\end{align}
which we need in order to carry out (4.5). Putting all this together yields
\begin{align}\label{4.10}
    \left(\frac{\partial\mu}{\partial m}\right)_{r,\lambda} = \Xi^{2}a^{2}\left(1+\frac{\Xi^{2}a^{2}\lambda\Delta_{\theta}}{2P^{2}}\right)I + \frac{3}{2}\Xi^{4}a^{4}\lambda J
\end{align}
with $I$ and $J$ defined in \eqref{4.6} and \eqref{4.9}. Since \eqref{4.10} is positive for $r>0$, this shows that $\mu$ is \textit{increasing} as the black hole mass is increasing. Thus, condition \eqref{4.1} is not satisfied by $\sin\theta$ or $\mu$ for $m>0$.

The last factor in \eqref{4.1} to check is $Q$. It is not difficult to show using  \eqref{PQdef}
that when the charge of the black hole vanishes, $Q>Q_{0}>0$, where $Q_{0} = Q(r,\lambda,m=0)$ \cite{al2019null}. Hence, $Q$ cannot reach zero faster for $m>0$ than it does in pure AdS, and therefore it does not go to zero for any $r>0$, when the black hole is uncharged. Thus, condition \eqref{4.1} is \textit{not} satisfied by the factor $Q$ in the Kerr-AdS case \cite{al2019null}. However, this changes once we impose charge on the black hole.

Figure \ref{fig:caustics_k(n)ads} is a plot of $Q(r)$ for both the Kerr-AdS and the Kerr-Newman-AdS black holes, for certain values of the parameters. As one can see in the 
KAdS case, $Q(r)$ 
is positive and never goes to zero for all $r>0$. However, in the KNAdS case it goes to zero inside the inner horizon of the black hole. Therefore, when $q\neq0$, i.e, when the black hole is charged, $Q(r)\rightarrow 0$ for $0<r<r_{-}$ where $r_{-}$ is the inner horizon radius of the Kerr-Newman-AdS black hole. Thus, we conclude that condition \eqref{4.1} \textit{is} satisfied for the KNAdS black hole and hence, caustics are formed between the ring singularity and the inner horizon of the KNAdS black hole. Condition \eqref{4.1} is not satisfied for any $r>r_{-}$, so caustics do not form anywhere outside the inner horizon of the KNAdS black hole. \par  Figure \ref{fig:caustics_l_infty} is a plot of $Q(r)$ in the limit where $l\rightarrow\infty$. Taking this limit of the KNAdS and KAdS metrics yields the well known, asymptotically flat, Kerr-Newman and Kerr metrics respectively. As one can see from the plot, caustics remain present for $r>0$ in the charged Kerr-Newman case and remain absent in the Kerr case, which is consistent with \cite{pretorius1998quasi}. 

\begin{figure}[t]
    \centering
    \includegraphics[scale=0.305]{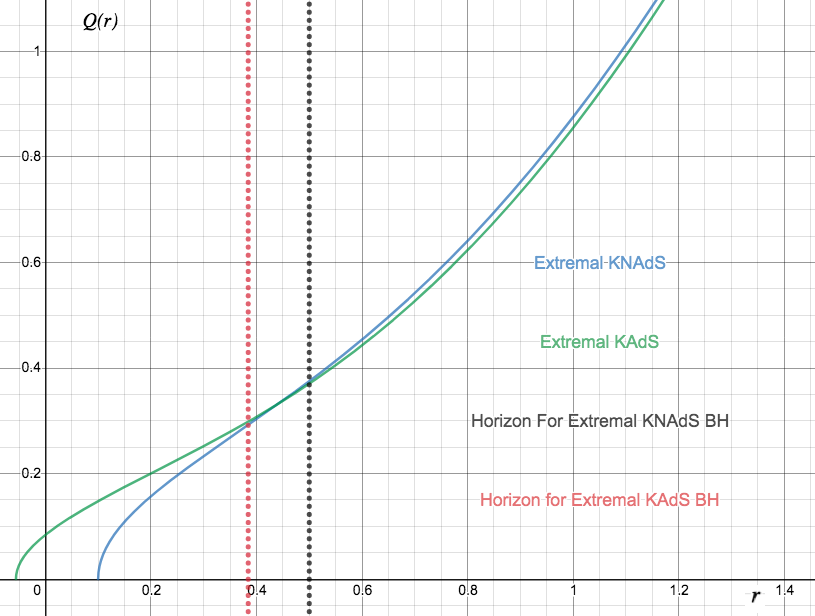}
    \caption{$Q(r)$ for the extremal KNAdS and KAdS black holes for $l=1$, $a=0.5$, $q=0.5$, and $\lambda = 0.8$. As one can see, $Q\to 0$ for $r>0$ in the extremal KNAdS case while $Q>0$ for all $r>0$ in the extremal KAdS case, indicating that caustics form for the charged black hole in the extremal limit.}
    \label{fig:caustics_ext}
\end{figure}

 By studying the positive roots of $\Delta_{r}$ from \eqref{KNAdS2}, it can be shown that there exists a critical mass for the KNAdS black hole, called the \textit{extremal} mass given by \cite{caldarelli2000thermodynamics}
\begin{widetext}
\begin{equation}\label{M_ext}
M_{ext} = \frac{l}{3\sqrt{6}}\left(\sqrt{\left(1+\frac{a^{2}}{l^{2}}\right)^{2} + \frac{12}{l^{2}}(a^{2}+q^{2})} + \frac{2a^{2}}{l^{2}} + 2\right)
\left(\sqrt{\left(1+\frac{a^{2}}{l^{2}}\right)^{2} + \frac{12}{l^{2}}(a^{2}+q^{2})} - \frac{a^{2}}{l^{2}} - 1\right)^{\frac{1}{2}}
\end{equation} 
\end{widetext}
The metric \eqref{KNAdS} describes a naked singularity for $M < M_{ext}$ and a black hole with an outer event horizon and an inner Cauchy horizon for $M > M_{ext}$. For $M=M_{ext}$, \eqref{KNAdS} describes an extremal black hole, where the event horizon and Cauchy horizon coincide. The extremal mass of the KAdS black hole is given by \eqref{M_ext} with $q=0$. Figure \ref{fig:caustics_ext} is a plot of the function $Q(r)$ for the extremal KNAdS and KAdS black holes. Once again, caustics appear in the charged case and are absent in the uncharged case.

\begin{figure}[t]
    \centering
    \includegraphics[scale=0.6]{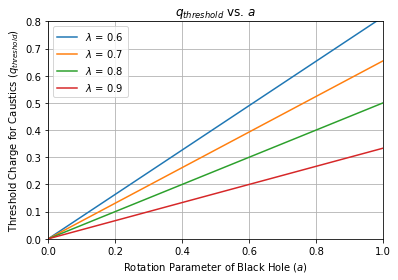}
    \caption{Threshold charge for caustics to form as a function of the rotation parameter of the black hole $a$, for constant values of $\lambda$. In this plot, $m=l=1$. $q_{threshold}$ grows linearly with $a$, indicating that the faster the black hole is spinning, the more charge it must have for caustics to form. This plot is the same in both the Kerr-Newman-AdS and Kerr-Newman cases.}
    \label{fig:threshold_q}
\end{figure}

Upon further examination, we find  that there is a threshold charge for caustics to develop in the KNAdS case as well as its asymptotically flat counterpart. The value of this threshold charge does not depend on the mass of the black hole nor the AdS length, $l$, so long as $a<l$ is satisfied. However, it does have a dependence on the rotation parameter, $a$, of the black hole. This is straightforwardly obtained by noting that for a caustic to form we must have $Q=0$ and $r>0$. In the limit that $r\to 0$ for vanishing $Q$ we find that 
\begin{equation}
    |q| > q_{threshold} =   \sqrt{\frac{1-\lambda}{\lambda}} a
\end{equation}
in order for a caustic to form. Figure \ref{fig:threshold_q} is a plot of the threshold charge for caustics to form, $q_{threshold}$, as a function of the rotation parameter of the black hole, $a$, holding $\lambda$ constant, for the KNAdS black hole. We see that $q_{threshold}$ grows linearly with the rotation of the black hole:   the faster the black hole is rotating, the more charge it must have  for caustics to form inside the inner horizon. This linear relationship between $q_{threshold}$ and $a$ holds true in the Kerr-Newman case as well.
 
%\\\\\\\ SECTION 5 \\\\\\\

\section{Ultra-Spinning Limit of $\text{KAdS}$ and $\text{KNAdS}$ Black Holes}

The ultra-spinning limit of the metric \eqref{KNAdS} is obtained by taking the $a\rightarrow l$ limit of \eqref{KNAdS} and (2.2) respectively. This effectively boosts the already rotating coordinate system to the speed of light and changes the nature of the spacetime qualitatively because it is no longer viable to return to a non-rotating reference frame once this limit is taken \cite{hennigar2015ultraspinning}. Taking this limit \textit{requires} the KNAdS geometry to be expressed in terms of a rotating coordinate system \cite{hennigar2015ultraspinning}. In order to avoid any coordinate singularities in the metric after taking the ultra-spinning limit, it is necessary to introduce a new coordinate given by $\psi = \phi/\Xi$. With this new coordinate, and taking the $a\rightarrow l$ limit, the metric \eqref{KNAdS} becomes
\begin{widetext}
\begin{equation} \label{5.1}
    ds^2\:=-\frac{\Delta}{\Sigma^2}\left[dt-l\sin^2\theta d\psi\right]^2\:+\frac{\Sigma^2}{\Delta}dr^2+\frac{\Sigma^2}{\sin^2\theta}d\theta^2 + \frac{\sin^4\theta}{\Sigma^2}\left[ldt-(r^2+l^2)d\psi\right]^2
\end{equation}
where
\begin{equation} \label{5.2}
    \Delta\:=\left(l+\frac{r^2}{l}\right)^2\:-2mr+q^{2}, \hspace{0.5cm}
    \Sigma^2=r^2+l^2\cos^2\theta
\end{equation}
\end{widetext}
It is important to notice that when the $a\rightarrow l$ limit is taken, the new coordinate $\psi$ becomes non-compact. The coordinate can be compactified via $\psi \sim \psi + \mu$, where $\mu$ is a dimensionless parameter which related to the chemical potential of the black hole. It can be straightforwardly shown that \eqref{5.1} is a solution of the Einstein-Maxwell equations for $\Lambda < 0$. Once again, the outer horizon is given by the largest root of $\Delta(r)$. It is also worth pointing out that \eqref{5.1} becomes singular at $\theta = 0, \pi$, however this part of the spacetime has been excised and represents some kind of boundary \cite{hennigar2015ultraspinning}. This indicates that topologically, this black hole is a sphere with punctures at both its poles and hence has a non-compact horizon. The metric \eqref{5.1} is an example of what is known as a \textit{Super-Entropic} black hole \cite{hennigar2015ultraspinning} insofar as it violates the \textit{Reverse Isoperimetric Inequality} \cite{cvetivc2011black}. In four dimensions, these black holes are also solutions to gauged supergravity \cite{gnecchi2014rotating}, which is highly important in the context of the AdS/CFT correspondence \cite{maldacena1999large}. Note that the ultraspinning KAdS black hole is also described by \eqref{5.1}, with $q=0$. \par
In order to study null hypersurfaces in \eqref{5.1}, it is necessary to define the ingoing/outgoing null generators in the same way as \eqref{uvdef}, with $r_{*} = r_{*}(r,\theta)$ once again. The PDE for $r_{*}(r,\theta)$, analogous to \eqref{rstareq} but in the ultra-spinning case, takes the form
\begin{equation} \label{5.3}
    \Delta\left(\partial_{r}r_{*}\right)^{2} + \sin^{2}\theta \left(\partial_{\theta}r_{*}\right)^{2} = \frac{\left(r^{2} + l^{2}\right)^{2}}{\Delta} - l^{2}
\end{equation}
where $\Delta$ is defined in \eqref{5.2}. To find a solution to this PDE, it is once again necessary to define two new functions
\begin{gather}
    Q^{2}(r) = \left(r^{2} + l^{2}\right)^{2} - l^{2}\lambda\Delta, \notag \\ P^{2}(\theta) = l^{2}\sin^{2}\theta\left(\lambda - 1\right)
\end{gather}
and a solution will be obtained by integrating the exact differential
\begin{equation} \label{5.5}
    dr_{*} = \frac{Q}{\Delta}dr + \frac{P}{\sin^{2}\theta}d\theta
\end{equation}
The first key difference between the KNAdS case and its ultra-spinning counterpart is the fact that from the definition of $P(\theta)$ in (5.4), one can see that $\lambda > 1$ is a requirement for solutions to be real. This constraint must hold for all values of $r$, in the function $Q$. Recall that $\lambda < 1$ was the assumption in the KNAdS case, so taking the ultra-spinning limit is forcing $\lambda > 1$. The most general solution to \eqref{5.5} is obtained by promoting $\lambda$ to be a function of $(r,\theta)$ and the procedure, as outlined in section two, is exactly the same with
\begin{equation} \label{5.6}
    F = \int_{r}^{r_{c}}\frac{dr^{\prime}}{Q} + \int_{0^{+}}^{\theta}\frac{d\theta^{\prime}}{P} + g^{\prime}(\lambda)
\end{equation}
where $r_{c}$ is some cut off distance needed because $Q$ is not defined for arbitrarily large values of $r$ due to the $\lambda > 1$ constraint. Using the inversion of differentials
\begin{gather}
    \Sigma^2 R^2 dr=\Delta Q\left(\sin^2\theta dr_{*} - P^2 \mu d\lambda\right) \notag \\
    \Sigma^2 R^2 d\theta=P\sin^2\theta\left(\Delta dr_{*} + Q^2 \mu d\lambda\right)
\end{gather}

\noindent which come from \eqref{mdl} and \eqref{5.5}, the metric \eqref{5.1} can be rewritten in terms of the coordinates $(t, r_{*},\theta,\lambda)$ as
\begin{gather}
    ds^{2} = \frac{\Delta\sin^{2}\theta}{R^{2}}\left(dr_{*}^{2} - dt^{2}\right) + R^{2}\sin^{2}\theta\left(d\psi - \omega_{US} dt\right)^{2}  \notag \\
   + \frac{\mu^{2}P^{2}Q^{2}}{R^{2}}d\lambda^{2}
\end{gather}
 where $\mu \equiv -\partial_{\lambda}F$ \footnote{It should be noted that this parameter is not in any way related to the chemical potential of the black hole mentioned previously} and
\begin{equation} \label{5.9}
    R^{2} \equiv \frac{g_{\psi\psi}}{\sin^{2}\theta} = \frac{l^{2}\sin^{2}\theta}{\Sigma^{2}}\left(2mr - q^{2}\right)
\end{equation}
\begin{equation} \label{5.10}
    \omega_{US} \equiv -\frac{g_{t\psi}}{g_{\psi\psi}} = \frac{l}{\Sigma^{2} R^{2}}\left[(r^{2} + l^{2})\sin^{2}\theta - \Delta\right]
\end{equation}
with $\omega_{US}$ being the ZAMO angular velocity of the ultra-spinning KNAdS black hole.  This entire formalism can easily be applied to the KAdS black hole, yielding the same equations but with $q=0$. 

\begin{figure}
    \centering
    \includegraphics[scale=0.27]{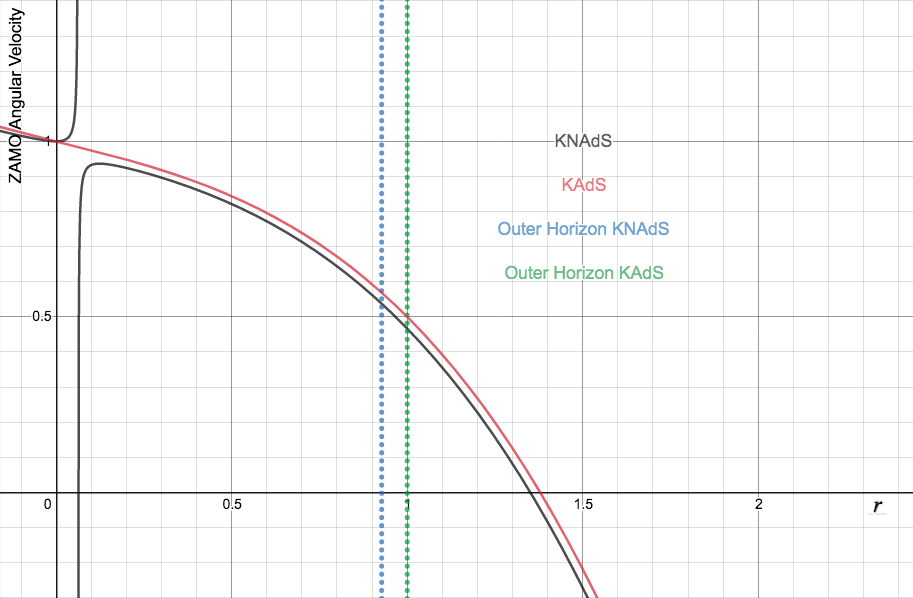}
    \caption{ZAMO angular velocity for the ultra-spinning KNAdS (black) and KAdS (red) black holes respectively with $m = 2$, $q = 0.5$, $l = 1$ and $\theta = \pi/2$. For large $r$, $|\omega_{US}|$ is slightly larger in the charged KNAdS case, than in the uncharged KAdS case.}
    \label{fig:zamo_US_lim}
\end{figure}

 Figure \ref{fig:zamo_US_lim} shows a plot of $\omega_{US}$ as a function of distance from the ultra-spinning KNAdS and KAdS black holes, respectively. The ZAMO angular velocity $\omega_{US}$ has the same peculiar feature of being positive for small values of $r$ and negative for large values of $r$ for both black holes, with $\omega_{US}$ diverging inside the charged black hole, as in the regular cases shown in figure 1. However, rather than asymptotically approaching a constant value as $r\rightarrow\infty$, it diverges to $-\infty$. This means that as an inertial observer gets farther away from the ultra-spinning KNAdS and KAdS black holes, the angular velocity of inertial frame dragging gets larger and larger. Here lies the second key difference between the regular KNAdS/KAdS cases and their ultra-spinning counterparts. This behaviour of $\omega_{US}$ will play an important role in understanding caustic formation in these spacetimes. 
 We easily obtain the induced metric on a null hypersurface embedded in the ultra-spinning KNAdS/KAdS spacetime from (5.8) by noting that for constant $v=t+r_{*}$ and $u = t-r_{*}$, $dv=du=0$. Therefore, $dt^{2} = dr_{*}^{2}$.
Thus, the induced metric is
\begin{equation} \label{5.11}
    dh_{US}^{2} =  R^{2}\sin^{2}\theta\left(d\psi - \omega_{US} dt\right)^{2} + \frac{\mu^{2}P^{2}Q^{2}}{R^{2}}d\lambda^{2}
\end{equation}
indicating that the null generators are rotating with the ZAMO angular velocity $\omega_{US}$.
By taking the determinant of \eqref{5.11}, the condition for the formation of caustics in the ultra-spinning case is
\begin{equation} \label{5.12}
    \mu PQ\sin\theta \rightarrow 0
\end{equation}
Once again, each factor in \eqref{5.12} needs to be studied individually to see where the condition satisfied. One can straightforwardly show that \eqref{5.12} is not satisfied by $\mu$, $P$ or $\sin\theta$, by similar arguments to the ones made in section four. As with the regular KNAdS case, condition \eqref{5.12} will be satisfied by the factor $Q$. \par
\begin{figure}
    \centering
    \includegraphics[scale=0.31]{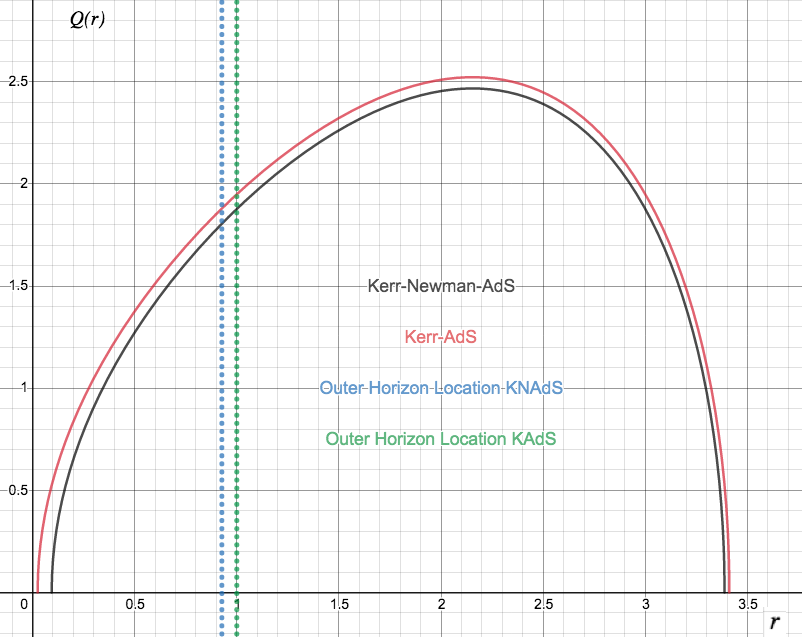}
    \caption{$Q(r)$ for the ultra-spinning KNAdS and KAdS geometries for $m=2$, $l=1$, $q=0.5$, $\lambda = 1.1$. The green dotted line is the outer horizon location of the uncharged Super-Entropic black hole, while the blue dotted line is the outer horizon location of the charged Super-Entropic black hole. Notice that caustics form closer to the black hole in the charged case.}
    \label{fig:caustics_US_lim}
\end{figure}
 Figure \ref{fig:caustics_US_lim} shows the function $Q(r)$ for the ultraspinning KNAdS and KAdS black holes respectively. We see that it goes to zero both inside and \textit{outside} the outer horizon of both black holes,   in strong contrast to its   KNAdS and KAdS counterparts \cite{al2019null}. This
 situation is rooted in the fact that we must have  $\lambda > 1$ for the ultra-spinning case. One physical explanation for the formation of caustics in these spacetimes is that the null generators are being dragged with the ZAMO angular velocity, $\omega_{US}$. Because this angular velocity is increasing as $r$ increases, there comes a point where the spacetimes are rotating so fast that the light cones fold up, and their volume goes to zero, signalling the formation of a caustic. Notice that caustics are forming closer to the ultra-spinning charged black hole than in the uncharged case. This is consistent with our physical explanation  -- from figure \ref{fig:zamo_US_lim} we see that spacetime is being dragged slightly faster, for a given value of large $r$, in the charged case than in the uncharged case;  caustics are thus expected to develop closer to the charged black hole. Caustics are also present inside both the charged and uncharged super-entropic black holes, forming closer to the outer horizon as the charge increases.
 
 Summarizing, taking the ultra-spinning limit of the KAdS black hole leads to caustic formation inside its inner horizon for arbitarily small values of $q$, including $q=0$. This
likewise has the effect of caustics forming outside of the horizon, with their formation occurring closer to the horizon as $q$ increases.
 
  Although caustics form at finite values of $r > r_{+}$, there is still a \textit{local} notion of causality given from the equivalence principle of General Relativity.

%\\\\\\\ SECTION 6 \\\\\\\

\section{Kruskal Extensions}

\subsection{Kruskal Coordinates for KNAdS Black Holes}
We start by noting that under the transformations 
\begin{equation} \label{6.1}
    -\frac{dU}{\kappa U} = du = d(t-r_{*}), \hspace{0.5cm} \frac{dV}{\kappa V} = dv = d(t+r_{*})
\end{equation} 
the KNAdS metric \eqref{3.9} becomes
\begin{gather} \label{6.2}
    ds^{2} = \frac{\Delta_{r}\Delta_{\theta} dUdV}{\kappa^{2}\Xi^{4}R^{2}UV} + R^{2}\sin^{2}\theta\left[d\phi - \frac{\omega}{2\kappa}\left(\frac{dV}{V} - \frac{dU}{U}\right)\right]^{2} \notag \\
    + \frac{P^{2}Q^{2}\mu^{2}}{\Xi^{4}R^{2}}d\lambda^{2}
\end{gather}
where 
\begin{equation} \label{6.3}
    \kappa_{\pm} = \frac{d\Delta_{r}}{dr}|_{r=r_{\pm}}
\vspace{0.1cm}
\end{equation}

\noindent is the surface gravity of the inner and outer horizons respectively, and $(U,V)$ are the Kruskal coordinates.\par
Looking at the metric (6.2), one can see that the first term is regular at the horizon surfaces $U=0$, $V=0$. The second term, however, is singular at these surfaces. In order to avoid this problem, it is necessary to define the advanced and retarded angular coordinates 
\begin{equation} \label{6.4}
    \varphi_{+} = \phi + \alpha(r,\lambda), \hspace{0.5cm} \varphi_{-} = \phi - \alpha(r, \lambda)
\end{equation}
where 
\begin{equation} \label{6.5}
    \alpha(r, \lambda) = a\Xi^{3}\left[\int_{0}^{r}\frac{r^{\prime 2} +a^{2}}{\Delta_{r} Q}dr^{\prime} - \int_{0}^{\theta}\frac{1}{\Delta_{\theta} P}d\theta^{\prime}\right]
\end{equation}
with $P$ and $Q$ defined in \eqref{PQdef}.
It is straightforward to show using \eqref{3.11}, \eqref{3.9} and \eqref{PQdef} that
\begin{equation} \label{6.6}
    d\alpha = \omega dr_{*} - Nd\lambda
\end{equation}
where 
\begin{widetext}
\begin{equation} \label{6.7}
    N = \frac{a\mu}{\Xi R^{2}\Sigma^{2}}\left[P^{2}(r^{2}+a^{2}) + Q^{2}\right] - \frac{a^{3}\Xi^{5}}{2}\left[\int_{0}^{r}\frac{r^{\prime 2} + a^{2}}{Q^{3}}dr^{\prime} + \int_{0}^{\theta}\frac{1}{P^{3}}d\theta^{\prime}\right]
\end{equation}

Therefore, \eqref{6.4} and \eqref{6.6} allow the metric (6.2) to be rewritten as
\begin{equation} \label{6.8}
    ds_{+}^{2} = \frac{\Delta_{r}\Delta_{\theta} dUdV}{\kappa^{2}\Xi^{4}R^{2}VU} + R^{2}\sin^{2}\theta\left(d\varphi_{+} - \omega dv + Nd\lambda\right)^{2} + \frac{\mu^{2}P^{2}Q^{2}}{\Xi^{4}R^{2}}d\lambda^{2}
\end{equation}
or
\begin{equation} \label{6.9}
     ds_{-}^{2} = \frac{\Delta_{r}\Delta_{\theta} dUdV}{\kappa^{2}\Xi^{4}R^{2}VU} + R^{2}\sin^{2}\theta\left(d\varphi_{-} + \omega dv - Nd\lambda\right)^{2} + \frac{\mu^{2}P^{2}Q^{2}}{\Xi^{4}R^{2}}d\lambda^{2}
\end{equation}
depending on the sheet of interest.
\end{widetext}

For example, \eqref{6.8} is regular on the future light sheets of both the outer and inner horizons of the KNAdS black hole and $\varphi_{+}$ is constant along each ingoing null generator. On the other hand, \eqref{6.9} is regular on the past light sheets of both the outer and inner horizons of the KNAdS black hole and $\varphi_{-}$ is constant along the outgoing null generators. In both of these cases, $N$ is regular. This is in agreement with what was done in \cite{pretorius1998quasi,al2019null}.

\subsection{Kruskal Coordinates for the Ultra-Spinning KNAdS Black Hole}
The method for constructing Kruskal coordinates for the ultra-spinning KNAdS black hole is similar to what was done above, with a few minor changes. We begin by using \eqref{6.1} to rewrite the metric (5.8) as
\begin{gather}
    ds^{2} = \frac{\Delta\sin^{2}\theta dUdV}{\kappa^{2}R^{2}UV} + R^{2}\sin^{2}\theta\left[d\psi - \frac{\omega_{US}}{2\kappa}\left(\frac{dV}{V} - \frac{dU}{U}\right)\right]^{2} \notag \\
    + \frac{\mu^{2}P^{2}Q^{2}}{R^{2}}d\lambda^{2}
\end{gather}
with $\omega_{US}$ being defined in \eqref{5.10}. The surface gravity $\kappa$, in the ultra-spinning case, is defined as
\begin{equation} \label{6.11}
    \kappa_{\pm} = \frac{d\Delta}{dr}|_{r=r_{\pm}}
\end{equation}
with $\Delta$ defined in \eqref{5.2}. Once again, the metric (6.10) is not regular at the horizon surfaces $U = 0$ and $V = 0$, due to the second term. Analogously to the above procedure, this is dealt with by defining the advanced and retarded compactified angular coordinates
\begin{equation} \label{6.12}
    \psi_{+} = \psi + \beta(r, \lambda), \hspace{0.5cm} \psi_{-} = \psi - \beta(r,\lambda)
\end{equation}
where
\begin{equation} \label{6.13}
    \beta(r, \lambda) = l\left[\int_{0}^{r}\frac{r^{\prime 2} + l^{2}}{\Delta(r^{\prime}) Q(r^{\prime},\lambda)}dr^{\prime} - \int_{0^{+}}^{\theta}\frac{1}{\sin^{2}\theta^{\prime}P(\theta^{\prime}, \lambda)}d\theta^{\prime}\right]
\end{equation}
with $P$ and $Q$ defined in (5.4). Making use of (5.7), (5.10) and (5.4), it is not hard to show that
\begin{equation} \label{6.14}
    d\beta = \omega_{US}dr_{*} - Nd\lambda
\end{equation}
where
\begin{widetext}
\begin{equation} \label{6.15}
    N \equiv \frac{l\mu}{\Sigma^{2}R^{2}}\left[P^{2}(r^{2}+l^{2}) + Q^{2}\right] - \frac{l^{3}}{2}\left[\int_{0}^{r}\frac{r^{\prime 2}+l^{2}}{Q^{3}}dr^{\prime} + \int_{0}^{\theta}\frac{1}{P^{3}}d\theta^{\prime}\right]
\end{equation}
Using \eqref{6.12} and \eqref{6.14}, we can write the metric (6.10) as
\begin{equation} \label{6.16}
    ds_{+}^{2} = \frac{\Delta\sin^{2}\theta dUdV}{\kappa^{2}R^{2}UV} + R^{2}\sin^{2}\theta(d\psi_{+} - \omega_{US}dv + Nd\lambda)^{2} + \frac{\mu^{2}P^{2}Q^{2}}{R^{2}}d\lambda^{2}
\end{equation}
or 
\begin{equation} \label{6.17}
    ds_{-}^{2} = \frac{\Delta\sin^{2}\theta dUdV}{\kappa^{2}R^{2}UV} + R^{2}\sin^{2}\theta(d\psi_{-} - \omega_{US}du - Nd\lambda)^{2} + \frac{\mu^{2}P^{2}Q^{2}}{R^{2}}d\lambda^{2}
\end{equation}
\end{widetext}

\noindent 
again, depending on which sheet is of interest.
\eqref{6.16} is regular for the future light sheets of both the inner and outer horizons of the ultra-spinning black hole, while \eqref{6.17} is regular on the past light sheets. Once again, $N$ is regular in both cases. This procedure can be straightforwardly applied to the ultra-spinning KAdS black hole and similar results are obtained.

%\\\\\\\ CONCLUSION \\\\\\\

\section{Conclusion}

We have inspected the light cone structure of the Kerr-Newman-AdS black hole by studying the properties of Null Hypersurfaces embedded in this spacetime. The equations that govern these properties were obtained through a three-dimensional lightlike foliation of the KNAdS spacetime. Caustics were found to develop for $0<r<r_{-}$, in strong contrast to what was found for the uncharged, Kerr-AdS case \cite{al2019null}. We then took the $l\to\infty$ limit of the KNAdS and KAdS spacetimes respectively and found that caustics still develop inside the asymptotically flat Kerr-Newman black hole, but remain absent in the Kerr geometry, consistent with \cite{pretorius1998quasi}. We also found a linear relation between the threshold value of charge needed for caustics to form and the rotation parameter of the black hole, indicating that the amount of electric charge the black hole must possess for the formation of caustics depends on how fast the black hole is spinning. As the black hole rotates faster, more charge is needed for caustics to form.

We have checked that our results are not a coordinate artifact by transforming our
metric to an asymptotically non-rotating coordinate system.  Equations \eqref{PQdef} remain the same up to a factor of $\Xi^2$ and all our results follow.

We then considered the ultra-spinning limit  of both the KNAdS and KAdS black holes, obtained by taking the $a\rightarrow l$ limit of the rotation parameter, yielding  examples of \textit{Super-Entropic} black holes \cite{hennigar2015ultraspinning}. Caustics appear inside the horizon for all possible values of $q$, including $q=0$.  Moreover,
caustics   develop at finite distances \textit{outside} both of these black hole geometries. This can be understood from an examination of  the ZAMO angular velocity, which gets larger as one moves away from the black hole. This means that there comes a point where the spacetime is rotating so fast that the light cones `fold up' and a caustic forms. 

We also derived the Kruskal extensions for both the KNAdS black hole and its ultra-spinning limit, based on the properties of null hypersurfaces we found. 

These results could be useful for studying rotating black holes in the AdS/CFT correspondence \cite{brown2016holographic, brown2015complexity} and in wave propagation in numerical relativity \cite{perlick2001living,winicour2005characteristic}.

%\abd{fix this reference because it is not the one for numerical relativity-see reference [2] for that}. %\abd{cite ref [9] from my paper here}. 
It would be interesting to study the light cone structure of more exotic types of AdS black holes. These include both charged and uncharged higher dimensional rotating AdS black holes, particularly those with multiple spins.
 
%\\\\\\\ Acknowledgements \\\\\\\

\begin{acknowledgments}
We thank Robie A. Hennigar for useful comments. This work was supported in part by the Natural Sciences and Engineering Research Council of Canada.
\end{acknowledgments}

\bibliographystyle{unsrt}
\bibliography{bibliography.bib}

\begin{thebibliography}{10}

\bibitem{pretorius1998quasi}
Frans Pretorius and Werner Israel.
\newblock Quasi-spherical light cones of the kerr geometry.
\newblock {\em Classical and quantum Gravity}, 15(8):2289, 1998.

\bibitem{al2019null}
Abdulrahim Al~Balushi and Robert~B Mann.
\newblock Null hypersurfaces in kerr--(a) ds spacetimes.
\newblock {\em Classical and Quantum Gravity}, 36(24):245017, 2019.

\bibitem{Carter:1968ks}
B.~Carter.
\newblock {Hamilton-Jacobi and Schrodinger separable solutions of Einstein's
  equations}.
\newblock {\em Commun. Math. Phys.}, 10(4):280--310, 1968.

\bibitem{carter1968hamilton}
Brandon Carter.
\newblock Hamilton-jacobi and schrodinger separable solutions of einstein’s
  equations.
\newblock {\em Communications in Mathematical Physics}, 10(4):280--310, 1968.

\bibitem{plebanski1976rotating}
Jerzy~F Plebanski and Marek Demianski.
\newblock Rotating, charged, and uniformly accelerating mass in general
  relativity.
\newblock {\em Annals of Physics}, 98(1):98--127, 1976.

\bibitem{hawking1999rotation}
Stephen~W Hawking, Christian~J Hunter, and Marika~M Taylor-Robinson.
\newblock Rotation and the ads-cft correspondence.
\newblock {\em Physical Review D}, 59(6):064005, 1999.

\bibitem{caldarelli2000thermodynamics}
Marco~M Caldarelli, Guido Cognola, and Dietmar Klemm.
\newblock Thermodynamics of kerr-newman-ads black holes and conformal field
  theories.
\newblock {\em Classical and Quantum Gravity}, 17(2):399, 2000.

\bibitem{Anabalon:2018qfv}
Andrés Anabalón, Finnian Gray, Ruth Gregory, David Kubiz\v~nák, and
  Robert~B. Mann.
\newblock {Thermodynamics of Charged, Rotating, and Accelerating Black Holes}.
\newblock {\em JHEP}, 04:096, 2019.

\bibitem{hennigar2015ultraspinning}
Robie~A Hennigar, David Kubiz{\v{n}}{\'a}k, Robert~B Mann, and Nathan Musoke.
\newblock Ultraspinning limits and super-entropic black holes.
\newblock {\em Journal of High Energy Physics}, 2015(6):96, 2015.

\bibitem{cvetivc2011black}
Mirjam Cveti{\v{c}}, GW~Gibbons, D~Kubiz{\v{n}}{\'a}k, and CN~Pope.
\newblock Black hole enthalpy and an entropy inequality for the thermodynamic
  volume.
\newblock {\em Physical Review D}, 84(2):024037, 2011.

\bibitem{gnecchi2014rotating}
Alessandra Gnecchi, Kiril Hristov, Dietmar Klemm, Chiara Toldo, and Owen
  Vaughan.
\newblock Rotating black holes in 4d gauged supergravity.
\newblock {\em Journal of High Energy Physics}, 2014(1):127, 2014.

\bibitem{maldacena1999large}
Juan Maldacena.
\newblock The large-n limit of superconformal field theories and supergravity.
\newblock {\em International journal of theoretical physics}, 38(4):1113--1133,
  1999.

\bibitem{brown2016holographic}
Adam~R Brown, Daniel~A Roberts, Leonard Susskind, Brian Swingle, and Ying Zhao.
\newblock Holographic complexity equals bulk action?
\newblock {\em Physical review letters}, 116(19):191301, 2016.

\bibitem{brown2015complexity}
Adam~R Brown, Daniel~A Roberts, Leonard Susskind, Brian Swingle, and Ying Zhao.
\newblock Complexity equals action.
\newblock {\em arXiv preprint arXiv:1509.07876}, 2015.

\bibitem{perlick2001living}
V~Perlick.
\newblock Living rev. rel. 7, 9 (2004).
\newblock {\em arXiv preprint arXiv:1010.3416}, 2001.

\bibitem{winicour2005characteristic}
J~Winicour.
\newblock Characteristic evolution and matching living rev.
\newblock {\em Relativity}, 8(10), 2005.

\end{thebibliography}
\end{document}